# ITERATIVE NETWORK-CHANNEL DECODING WITH COOPERATIVE SPACE-TIME TRANSMISSION

Saikat Majumder and Shrish Verma

Department of Electronics and Telecommunication Engineering,
National Institute of Technology Raipur, India

## ABSTRACT

*One of the most efficient methods of exploiting space diversity for portable wireless devices is cooperative communication utilizing space-time block codes. In cooperative communication, users besides communicating their own information, also relay the information of other users. In this paper we investigate a scheme where cooperation is achieved using two methods, namely, distributed space-time coding and network coding. Two cooperating users utilize Alamouti space time code for inter-user cooperation and in addition utilize a third relay which performs network coding. The third relay does not have any of its information to be sent. In this paper we propose a scheme utilizing convolutional code based network coding, instead of conventional XOR based network code and utilize iterative joint network-channel decoder for efficient decoding. Extrinsic information transfer (EXIT) chart analysis is performed to investigate the convergence property of the proposed decoder.*

## 1. INTRODUCTION

Wireless adhoc and sensor networks having a large number of low power wireless nodes have attracted a lot of attention from the researchers recently. The main challenge of wireless sensor networks is to achieve proper balance between transmit/processing power and quality of service. However, such multi-terminal systems are limited by impairments due to wireless channels, such as fading, and interference. Such low power portable devices are further constrained by limited computational capabilities and power consumption due to computationally complex algorithms. This limitation due to low computational capabilities of sensor nodes can be addressed by utilizing modern development in power efficient microelectronic devices or by shifting the computational load to the base station. The later technique involves designing systems with low encoder complexity and relatively computationally intensive decoder at the base-station.

On the other hand, due to fading the transmission over wireless channels suffer from severe time varying attenuation in signal strength. For a point to point wireless communication system, effect of fading is mitigated using multiple antennas at transmitter and receiver. Since, wireless sensor nodes are too small to accommodate multiple antennas on a single terminal, several nodes can cooperate to form a virtual multiple-input multiple output (MIMO) system [1-2]. Cooperative communication has emerged as an accepted method for achieving transmit diversity for mitigation of fading effect at the receivers. In cooperative communication transmitting users use one another's antenna to realize the benefit of MIMO. There are many cooperative strategies to achieve efficient node cooperation, such as amplify and forward (AF) [2,3,4], decode and forward (DF) [5], and coded cooperation [6,7]. In AF protocol, relay nodes retransmit amplified versions of the signal received from source. Amplification coefficients at the relay nodes control the performance at the destination. Whereas in DF, relay nodes first detect the received symbol using hard decision and then forwards the reencoded signal to destination. Coded cooperation achieves space diversity by forwarding different segment of a channel code through different paths. Various improvements in these fundamental techniques has been proposed in recent years. An AF technique is proposed in [8] where the expected distortion performances with progressive transmission and superposition coding is investigated. Zhou et. al [7] proposed a distributed joint source-channel coding technique that exploits source relay correlation.

Besides distributed MIMO techniques, network coding [9] has also emerged as preferred method for obtaining cooperative diversity. A simple model where diversity can be obtained with network coding is multiple access relay channel (MARC). A simple such model consists of two cooperating user nodes and an intermediate relay with performs network coding [10] and decoding is performed at the base station on the principles of turbo code. Ahsin and Slimane [11] proposed a similar scheme for MARC using the principles of product codes. Authors in [12,13] proposed schemes which combines the benefits of space-time codes and network coding for cooperative communication. The authors use simple XOR based network coding at the relay for obtaining diversity. They have demonstrated that combination of Alamouti space-time code and network coding outperforms system based only on Alamouti coded cooperation.

In this paper, we improve upon the research in [12] by application of the concept of product code and iterative network-channel decoding. The proposed scheme uses punctured convolutional code as network code at the relay and Reed-Solomon code as channel code. The class of Reed-Solomon error correction codes is well known in technical literature and is adopted in many communication protocols. Our main innovation is iterative network-channel decoding at the receiver using the principles derived from soft decoding of concatenated Reed-Solomon convolutional codes [14]. The proposed algorithm enables network code and channel decoders to exchange soft information iteratively and may yield a capacity approaching performance. We apply extrinsic information obtained from soft network decoder to soft-input soft-out (SISO) decoder for Reed-Solomon code [19,15] through an interleaver. The extrinsic output of SISO Reed-Solomon decoder is applied back to network-decoder. SISO decoding of Reed-Solomon code allows its decoding beyond minimum separable distance (MDS) capability, in contrast to popular approach of hard decision decoding of Reed-Solomon code. An extrinsic information transfer (EXIT) [21] characteristics of the proposed algorithm is presented, leading to the insights of its iterative decoding behaviour and design criteria for network and channel code.

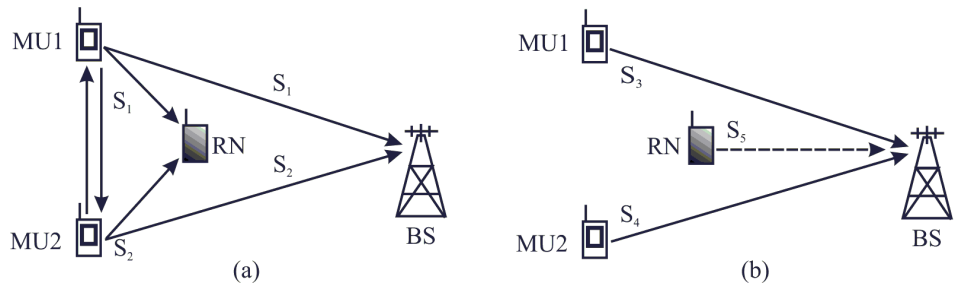

Figure 1. Space-time and network coded cooperation. (a) In first phase, user nodes broadcast information to relay, base station and other user node. (b) Second phase consists of retransmission by user nodes and relay node.

Rest of the article is organized as follows. Section 2 describes the system model under consideration. Cooperative space-time transmission, network coding and the proposed iterative network-channel decoder is discussed in this section. Section 3 analyzes the decoder using EXIT chart for fading channel. Simulation results are given in section 5. Finally, section 6 provides the concluding remarks and suggestion for future work.

## 2. SYSTEM MODEL

We consider the scenario shown in Figure 1, with two mobile users (MU1 and MU2) nodes communicating information to a common base station (BS). The two MU nodes cooperate with each other using Alamouti space-time block code. In addition, a dedicated relay node performs network coding on the information received from both the information nodes. We further assume that all the nodes are using orthogonal channels with relay node operating in half-duplex mode. Similar scenario is presented in [12] for evaluating the performance improvement due to additional network coding node. But instead of simple XOR based network coding, we consider convolutional encoder as network code and evaluate the performance of the system in iterative decoding scenario. Next we describe the channel coding operation at the MU nodes and relay.

### 2.1. Encoding at Mobile User Nodes

The bits from each source is grouped into *m* bit symbols belonging to Galois field GF(*N*+1), with $N = 2^m - 1$. The encoding at MU nodes is shown in Figure 2(a). The symbols are coded with (*N*, *K*) Reed-Solomon code, where *K* is the number of information symbols in a codeword. The Reed-Solomon code has dual functions; first, it is efficient against burst errors, since a sequence of $m + 1$ consecutive bit errors can affect at most two code symbols. Second, Reed-Solomon code aids in iterative joint network-channel decoding as discussed in next sections. In time slot 1 and 2, $L_i$ codewords are generated (each codeword consists of *N* symbols), where $i = 1, 2$ indicates the MU. They are grouped into a frame and interleaved with $\Pi_i$. The stream is then formed into matrix of size $L_i \times N$, where each row forms a packet. The symbols are translated into bits, modulated and broadcast to BS, relay and other MU.

### 2.2. Network Coding at Relay

Figure 2(b) shows the encoding operation at the relay node. The relay node overhears transmission from both the MU nodes during time slot $t = 1, 2$, decodes and reencodes them. The reencoded packets are ordered into matrix of size $(L_1 + L_2) \times mN$ bits. The rows from the two cooperating sources are arranged alternately as shown in Figure 3. In this research we use recursive systematic convolutional code (RSCC) as network code [11] instead of XOR based network code. RSCC of rate $(L_1 + L_2)/N_N$ is applied on each column of $(L_1 + L_2)$ bits, and parity bits are obtained. Puncturing may be applied on the parity bits to attain necessary code rates. Network code is obtained from these parity check bits and each row is transmitted as packet to the BS. Thus each row encounters different channel and bits in a row suffer from same amount of fading.

### 2.3. Cooperative Communication Protocol

The transmission of the message is accomplished in two phase or five time slots. In the first phase (Figure 1(a)), the mobile users MU1 and MU2 broadcast their messages $S_1$ and $S_2$ over wireless channel, respectively. This being broadcast phase, the transmitted messages are received by the BS, relay and the other user.

Table I

| Time Slot (*t*) | 1 | 2 | 3 | 4 | 5 |
|---|---|---|---|---|---|
| Transmitter | MS1 | MS2 | MS1 | MS2 | RN |
| Message | $S_1$ | $S_2$ | $S_3$ | $S_4$ | $S_5$ |

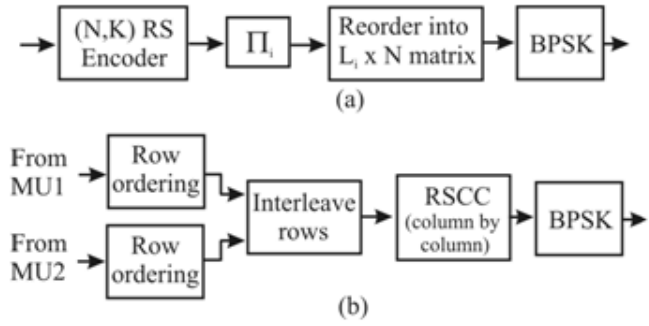

Figure 2. (a) Block diagram of encoder at mobile user nodes, (b) Network coding operation at relay node.

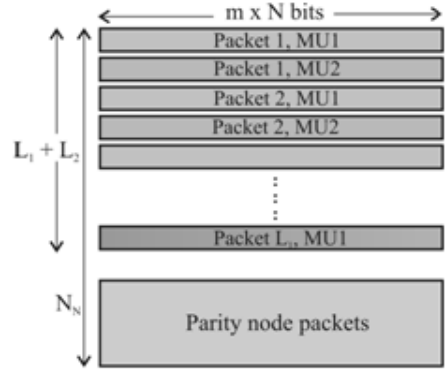

Figure 3. Data packets from two sources are ordered in alternate rows. Total number of rows from two sources is L1+L2. Remaining rows of parity check bits are calculated and transmitted from the relay node.

In the next multiple access phase, the cooperating users MU1 and MU2, using Alamouti STBC scheme, transmit $S_3 = -S_2^*$ and $S_4 = S_1^*$, respectively to the BS. The relay node decodes the signal received from both the users, reencodes them using Reed-Solomon code. Bits from both the sources are combined, network coded and transmitted to the destination. The scenario in second phase is illustrated in Figure 1(b). As mentioned earlier, all the nodes transmit in orthogonal channels (separate time slots), the Table 1 details the channel assignment for different transmitting nodes.

## 2.4. Iterative Network-Channel Decoder

The signal received at the BS after two phases of transmission and cooperative Alamouti relaying from two MS, in matrix vector notation, is

$$\begin{bmatrix} y_1 \\ y_2 \end{bmatrix} = \begin{bmatrix} h_1 & h_2 \\ h_2^* & -h_1^* \end{bmatrix} \begin{bmatrix} s_1 \\ s_2 \end{bmatrix} + \begin{bmatrix} n_1 \\ n_2 \end{bmatrix} \quad (1)$$

where, $h_i$, $i = 1, 2$ denotes complex multiplicative fading coefficients for MU-BS channel with $E\{|h_i|^2\} = 1$ and is assumed to be constant for at least one codeword duration. It is worth mentioning again that each row of bits of matrix in Figure 3 undergoes different fading. Utilizing (1), log-liklihood ratios (LLR) of the received packets from the two sources are obtained as $\Gamma_1$, $\Gamma_2$, respectively, using soft-output Alamouti decoder [17]. On the other hand, signal received from the relay node is

$$y_3 = h_3 s_3 + n_3 \quad (2)$$

and the corresponding LLR is calculated as $\Gamma_3 = (|y_3 + h_3|^2 - |y_3 - h_3|^2)/N_0$. Channel state information $h_i$ is assumed to be available at the base station. The rows of LLR $\Gamma_i$, i = 1, 2, 3 are

stacked over one another to form the matrix $\Gamma^{ch}$ in the order given in Figure 3. Output $\Gamma^{ch}$ of soft Alamouti decoder consists of alternate rows of LLRs of bits of Reed-Solomon codes from MU1 and MU2, while last $(N_N - L_1 - L_2)$ rows are LLR of bits received form RN.

At the receiver RSCC and Reed-Solomon code can be considered a concatenated code structure and can be decoded iteratively [14]. The next stages consists of iterative soft decoding process in which BCJR algorithm [18] is applied along the columns for soft decoding of RSCC and Jiang-Narayanan (JN) algorithm [19] for iterative soft decoding of Reed-Solomon codes along rows. JN algorithm or adaptive belief propagation (ABP) is a significant departure from the traditional hard decision decoding of Reed-Solomon codes. This algorithm operates in three stages. In the first stage the parity check matrix is adapted according to the incoming LLR, and in the second stage sum-product algorithm [20] is applied to calculate the extrinsic information. After finite number of iterations, Berlekemp-Messy algorithm for decoding Reed-Solomon code is applied to the hard-decisions made on the updated LLR. Besides being able to decode errors beyond maximum distance separable (MDS) capability, JN algorithm enables iterative soft decision decoding in conjunction with other soft decision decoders and equalizers. Extrinsic information is passed between the two SISO decoders for finite number of iterations or until decoding of all the Reed-Solomon codes is successful. The block diagram of the proposed decoder is shown in Figure 4.

An iteration of BCJR algorithm on the columns of LLR matrix $\Gamma^{ch}$ generates extrinsic LLR denoted by $\Gamma^e$. The LLR matrix $\Gamma^e$ consists of alternate rows of extrinsic LLR for the two sources, which are isolated at next stage into $\Gamma_1^e$ and $\Gamma_2^e$. Applying deinterleaving mapping $\Pi_i^{-1}$ on extrinsic LLR $\Gamma_i^e$, a priori LLR $L_i^a$ for next stage of iterative decoder is obtained. An iteration of JN algorithm is applied on all the rows of $L_i^a$ independently and resulting extrinsic information is saved as $L_i^e$. If a row of Reed-Solomon code satisfies the parity check requirement, it is decoded and saved in $\bar{S}_i$. Extrinsic information from both the RS decoders is interleaved, combined (into the matrix in Figure 3) and applied as a priori information to the BCJR decoder. This constitutes an iteration of the proposed decoder.

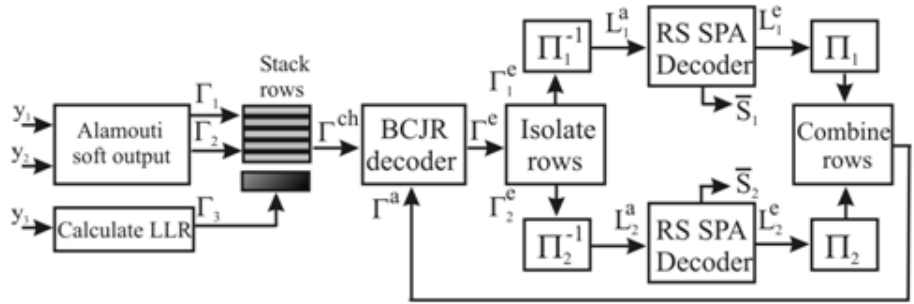

Figure 4. Block diagram of the proposed decoder

## 3. EXIT CHART ANALYSIS

Extrinsic information transfer (EXIT) chart has emerged as a successful method for predicting the convergence behaviour of various concatenated and iterative systems. In this section, EXIT chart is used to analyze the iterative decoding behaviour of the proposed scheme. The fundamental assumption of EXIT chart is that extrinsic information passed from one SISO decoder to other is a Gaussian random variable. The LLR *a* of a priori input for uncoded information *s* is modelled as

$$a = \mu_a x + n_a \qquad (3)$$

where $x$ is binary antipodal form of information symbols, $n_a$ is a Gaussian random variable with zero mean and variance $\sigma_a^2$. The variance must satisfy the condition $\mu_a = \sigma_a^2/2$. The mutual information between $a$ and $x$ is defined as

$$I(x,a) = \frac{1}{2} \sum_{x=\pm 1} \int_{-\infty}^{+\infty} f_a(\xi|x) \log_2 \frac{2 f_a(\xi|x)}{f_a(\xi|x=-1) f_a(\xi|x=+1)} d\xi \qquad (4)$$

where $f_a(\xi|x)$ is conditional probability density function associated with a priori LLR $a$. Therefore, for a priori LLR $a$, mutual information is given as $I_a = I(s;a)$. Similarly, mutual information for extrinsic output $e$ is obtained as $I_e = I(s;e)$. To obtain EXIT chart, for given values of $I_a = (0,1)$ we artificially generate the a priori inputs $a$, which are fed to SISO module. Then the corresponding decoding algorithm of the block is invoked to produce extrinsic output $e$. The mutual information $I_e$ is then evaluated using relation (4). Finally, EXIT chart is obtained as the graphical plot between $I_a$ and $I_e$. For decoding without any residual error, $I_e$ should equal 1 for some value of $I_a$.

Figure 5(a) shows the EXIT characteristics of the proposed decoder with (31,25) Reed-Solomon code. The inner decoder (decoder 1) consists of the cooperative Alamouti space-time decoder and punctured $(7,5)_8$ convolutional code acting as network code with overall code rate of 2/5. The $(I_a, I_e)$ curves are plotted with inner decoder 1 for average channel $E_b/N_0$ of 2 dB and 3 dB. Inverse EXIT characteristics $(I_e, I_a)$ of decoder 2 (outer RS decoder consisting of JN algorithm) is also shown in the figure. It shows, at 6 dB, the tunnel starts to open between EXIT curve of decoder 1 and decoder 2, and at 7 dB, the tunnel is completely open. Therefore, the decoder bit error rate (BER) cliff is expected to start at 6 dB and can be verified in Figure 7. Similarly, EXIT chart for the proposed system with (15,7) is given in Figure 5(b), where turbo-cliff starts at 5 dB.

## 4. SIMULATION RESULTS

In this section we demonstrate through simulations that cooperative space-time coded iterative network-channel decoder outperforms system utilizing XOR based network coding. The performance is evaluated for BPSK modulated signal transmitted over Rayleigh block fading channel, i.e. the channel fading coefficient is assumed to be constant for the duration of one codeword. As explained earlier, we consider MARC scheme with two users cooperatively transmitting using Alamouti STBC to the BS. An intermediate relay node assists in the transmission through network coding. The SNR of MU-BS and RN-BS is assumed to be same unless mentioned otherwise.

First we investigate the iterative convergence behavior of the proposed design. Figure 6 shows BER performance of the proposed network-channel decoder with (31, 25) Reed-Solomon code as component. It can be observed that iterative decoding gain is obtained for $E_b/N_0 \geq 6$ dB, as predicted in EXIT chart of Figure 5(a). Error rate decreases with increase in iterations and there is no significant improvement in BER after 20 iterations. Figure 7 shows the performance of the iterative decoder with (15,7) Reed-Solomon code, in which iteration gain starts for $E_b/N_0 \geq 5$ dB. Best BER performance is achieved for 5 iterations or more. This corresponds to EXIT chart in Figure 5(b).

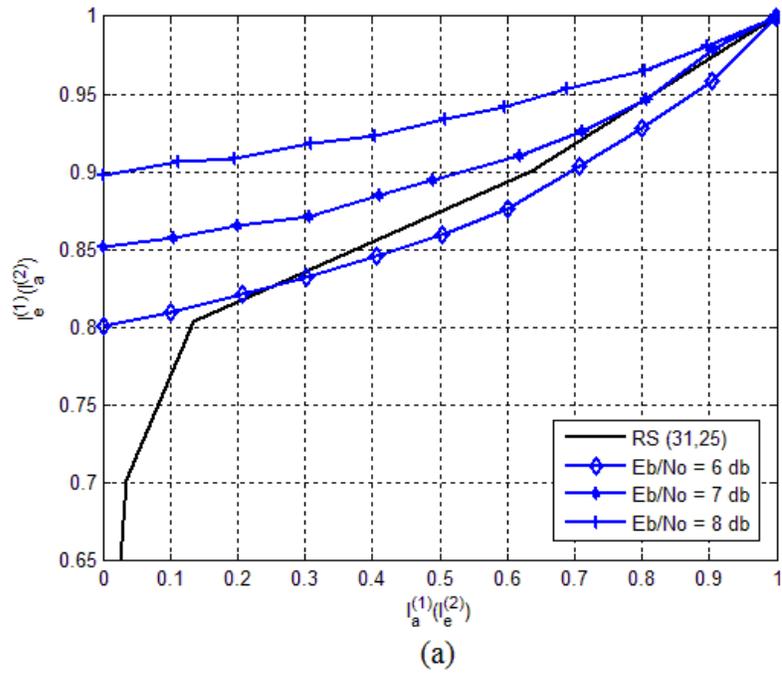

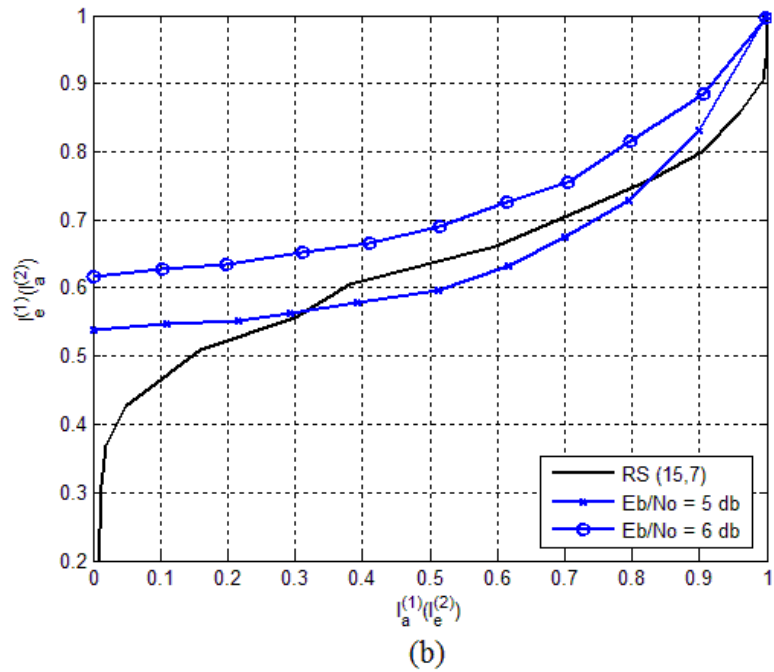

Figure 5. EXIT chart of the proposed iterative soft decoder utilizing (a) (31,25) and (b) (15,7) Reed-Solomon code as decoder 2.

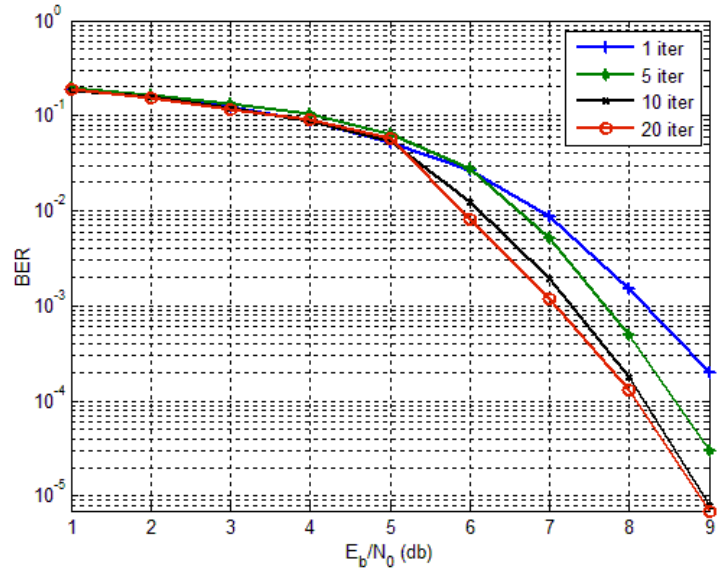

Figure 6. BER performance of the proposed scheme using (31,25) Reed-Solomon code with number of decoding iterations over Rayleigh block fading channel.

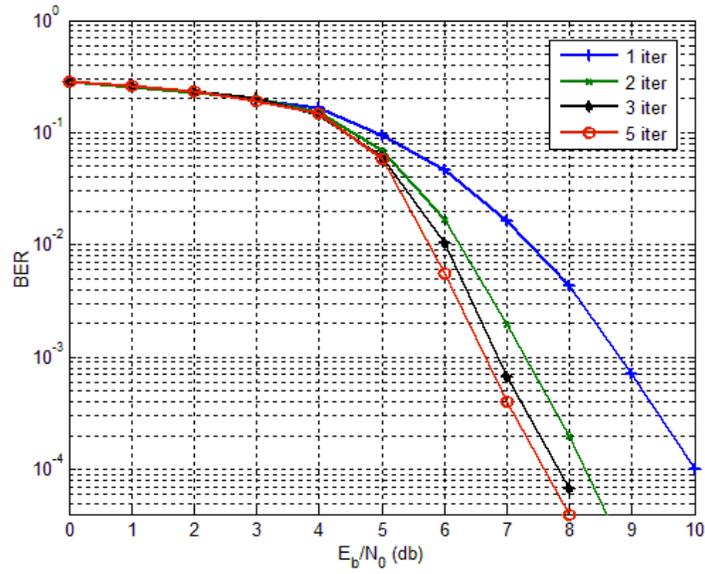

Figure 7. BER performance of the proposed scheme using (15,7) Reed-Solomon code with number of decoding iterations over Rayleigh block fading channel.

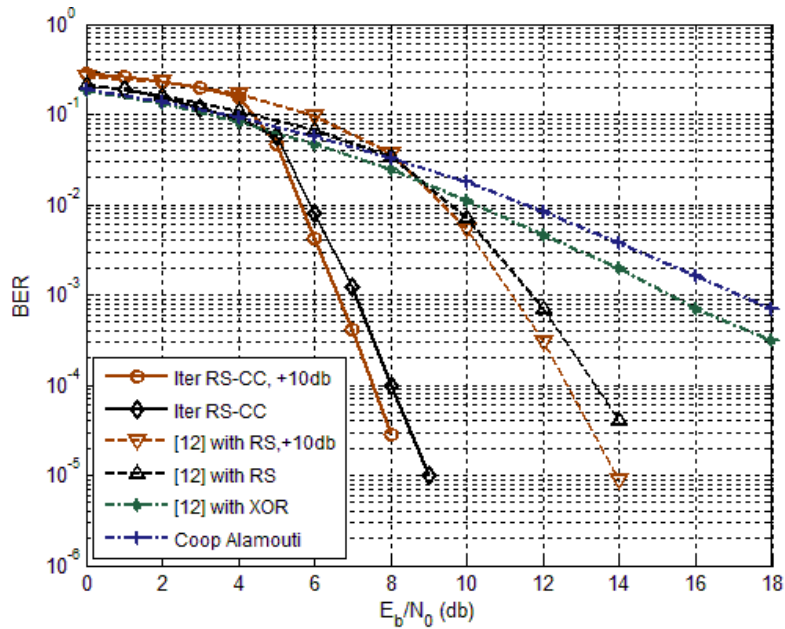

Figure 8. Performance comparison of the proposed network-channel iterative decoder with a reference scheme employing XOR based network code. (31,25) Reed-Solomon code is used as channel code in both cases for fair comparison. BER performance are also compared for the case when SNR at relay is 10 dB higher than direct links. BER curves of the reference system without using Reed-Solomon code is also shown.

Next we compare the BER performance of the proposed system to baseline scheme in [12] which applies XOR based network coding. Figure 8 illustrates the BER curves of proposed scheme with (31,25) Reed-Solomon codes as channel encoder. The network code is derived by puncturing rate ½ RSCC of generator polynomial $(5,7)_8$ to rate 2/3 code. The result is compared to BER curve obtained with baseline system utilizing (31,25) Reed-Solomon code as channel code at the MU nodes and XOR based network coding at RN. Overall code rate in both the cases is 2/5. It can be observed that proposed scheme benefits from iterative decoding of network and channel codes, where there is an improvement of 5 dB in $E_b/N_0$ for achieving BER of $10^{-4}$. Performance of all the schemes can be improved if the link between relay node and the base station is better. Figure 6 also illustrates BER for the all the above cases with SNR of relay-base station link 10 dB better than direct link. Performance of the baseline scheme without RS coding at the source is also shown for reference. Stronger RSCC used as network code can improve the error performance of the scheme as illustrated in Figure 9.

## 5. CONCLUSION

In this paper we proposed a novel iterative decoder for space-time-network coded cooperation over wireless network. With Reed-Solomon code as channel code and convolutional code based network coding and transmitted with space-time cooperation, the receiver is a cascade of SISO modules of BCJR decoder and ABP algorithm. Significant improvement in error rate performance is obtained with proposed scheme compared to scheme using XOR based network code. The EXIT analysis of the proposed decoder is presented to validate the decoding convergence results. Future work could be to reduce the complexity of the iterative decoder by using lesser complex algorithm compared to ABP algorithm.

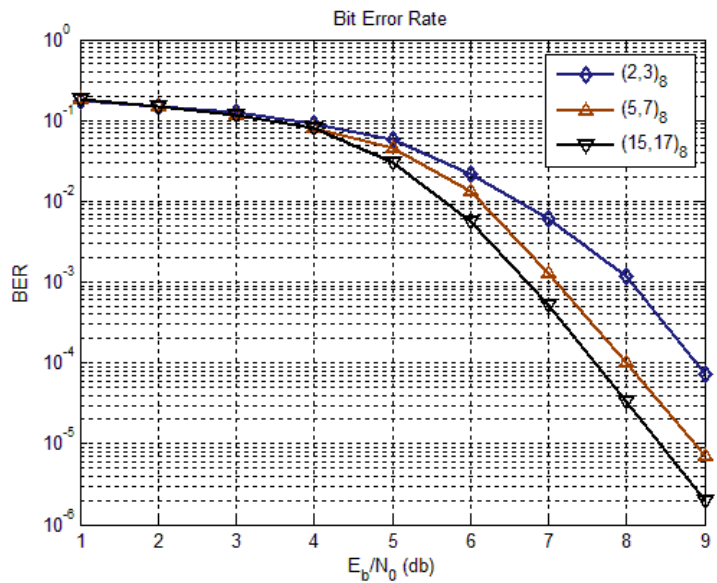

Figure 9. Performance of the proposed iterative scheme using (31,25) Reed-Solomon code for different RSCC used as network code.

## REFERENCES

[1] Yindi Jing, Hassibi, B. (2006) Distributed Space-Time Coding in Wireless Relay Networks. IEEE Transactions on Wireless communications, 5(12), pp.3524-3536. doi: 10.1109/TWC.2006.256975

[2] Laneman, J.N., Wornell, G. W. (2003) Distributed space-time-coded protocols for exploiting cooperative diversity in wireless networks, IEEE Transactions on Information Theory, 49(10), pp.2415-2425. doi: 10.1109/TIT.2003.817829

[3] Yi Zhao, Adve, R., Teng Joon Lim (2006) Improving Amplify-and-Forward Relay Networks: Optimal Power Allocation versus Selection. Proceedings of the 2006 IEEE International Symposium on Information Theory, 9-14 July 2006, pp.1234-1238.

[4] Sheng Yang, Belfiore, J.-C. (2007) Towards the Optimal Amplify-and-Forward Cooperative Diversity Scheme," IEEE Transactions on Information Theory, 53(9), pp.3114-3126. doi: 10.1109/TIT.2007.903133

[5] Zhao, B., Valenti, M.C. (2003) Distributed turbo coded diversity for relay channel. Electronics Letters , 39(10), pp.786-787. doi: 10.1049/el:20030526

[6] Hunter, T.E., Nosratinia, A. (2006) Diversity through coded cooperation. IEEE Transactions on Wireless communications, 5(2), pp.283-289. doi: 10.1109/TWC.2006.1611050

[7] Xiaobo Zhou, Meng Cheng, Anwar, K., and Matsumoto, T. (2012) Distributed joint source-channel coding for relay systems exploiting source-relay correlation and source memory. EURASIP Journal on Wireless Communications and Networking, 2012:260. doi:10.1186/1687-1499-2012-260

[8] Sethakaset, U., Quek, T.Q.S., Sumei Sun (2011) Joint Source-Channel Optimization over Wireless Relay Networks. IEEE Transactions on Communications, 59(4), pp.1114-1122. doi: 10.1109/TCOMM.2011.012711.090614

[9] Yingda Chen, Kishore, S., Jing Li (2006) Wireless diversity through network coding. Proceedings of the Wireless Communications and Networking Conference, (WCNC 2006). pp.1681-1686. doi: 10.1109/WCNC.2006.1696541

[10] Hausl, C., Dupraz, P. (2006) Joint Network-Channel Coding for the Multiple-Access Relay Channel, Proceedings of the Sensor and Ad Hoc Communications and Networks (SECON '06). pp.817-822, 28-28 Sept. 2006. doi: 10.1109/SAHCN.2006.288566

[11] Ahsin, Tafzeel ur Rehman and Slimane, Slimane Ben (2012) A Joint Channel-network Coding Based on Product Codes for the Multiple-access Relay Channel. ISRN Communications and Networking, 2012. doi: 10.5402/2012/837815

[12] Menghwar, G. D. et al. (2012) Cooperative space-time codes with network coding. EURASIP Journal on Wireless Communications and Networking, 2012(1). doi: 10.1186/1687-1499-2012-205
[13] Menghwar, G.D., Shah, A.A., Mecklenbrauker, C.F. (2009) Cooperative space-time codes with opportunistic network coding with increasing numbers of nodes. Proceedings of the 6th International Symposium on Wireless Communication Systems (ISWCS 2009), pp.536-539, 7-10 Sept. 2009. doi: 10.1109/ISWCS.2009.5285254
[14] Chen, L. (2013). Iterative Soft Decoding of Reed-Solomon Convolutional Concatenated Codes. IEEE Transactions on Communications, 61(10), pp.4076-4085. doi: 10.1109/TCOMM.2013.082813.120943
[15] Bellorado, J., Kavčić, A., Marrow, M., Li Ping (2010) Low-Complexity Soft-Decoding Algorithms for Reed–Solomon Codes—Part II: Soft-Input Soft-Output Iterative Decoding. IEEE Transactions on Information Theory, 56(3), pp.960-967. doi: 10.1109/TIT.2009.2039091
[16] Maunder, R. G., Wang, J., Yang, L. L. (2010). Near-capacity variable-length coding: regular and EXIT-chart-aided irregular designs (Vol. 20). John Wiley & Sons.
[17] Cho, Y. S., Kim, J., Yang, W. Y., Kang, C. G. (2010). MIMO-OFDM wireless communications with MATLAB. John Wiley & Sons.
[18] Bahl, L., Cocke, J., Jelinek, F., Raviv, J. (1974) Optimal decoding of linear codes for minimizing symbol error rate. IEEE Transactions on Information Theory, 20, (2), pp.284-287. doi: 10.1109/TIT.1974.1055186
[19] Jiang, Jing, Narayanan, K.R. (2006) Iterative Soft-Input Soft-Output Decoding of Reed-Solomon Codes by Adapting the Parity-Check Matrix. IEEE Transactions on Information Theory, 52(8), pp.3746-3756. doi: 10.1109/TIT.2006.878176
[20] Costello, D., & Lin, S. (2004). Error control coding. Pearson Higher Education.
[21] Ten Brink, S. (2001). Convergence behavior of iteratively decoded parallel concatenated codes. IEEE Transactions on Communications, 49(10), 1727-1737.